\documentclass[english,pra,twocolumn,groupedaddress,reprint]{revtex4}
\usepackage{graphicx}
\usepackage{amsfonts}
\usepackage[T1]{fontenc}
\usepackage[latin9]{inputenc}
\usepackage{color}
\usepackage{amsmath}
\usepackage{amssymb}
\usepackage{esint}
\usepackage{comment}
\usepackage{bbold}
\usepackage{tikz}
\usepackage{minitoc}
\usepackage{bm}

\begin{document}

\title{Room-temperature steady-state entanglement in a four-mode optomechanical system}
\author{Tao Wang\footnote{suiyueqiaoqiao@163.com (taowang@thnu.edu.cn)} $^{1,2}$, Rui Zhang $^{1}$ and Xue-mei Su\footnote{suxm@jlu.edu.cn} $^{1}$}
\affiliation{$^{1}$College of Physics, Jilin University, Changchun 130012, People's Republic of China}
\affiliation{$^{2}$College of Physics, Tonghua Normal University, Tonghua 134000, People's Republic of China}

\begin{abstract}
Stationary entanglement in a four-mode optomechanical system, especially under room-temperature, is discussed. In this scheme, when the coupling strengths between the two target modes and the mechanical resonator are equal, the results cannot be explained by the Bogoliubov-mode-based scheme. This is related to the idea of quantum-mechanics-free subspace, which plays an important role when the thermal noise of the mechanical modes is considered. Significantly prominent steady-state entanglement can be available under room-temperature.~~~~\newline
~~~~\newline
\end{abstract}

\maketitle

\section{Introduction}

Entanglement is a key resource in quantum information processing \cite{Horodecki09}, which has been intensively investigated in microscopic systems, such as cavity-QED \cite{Haroche01,guo00,song1,song2,song3}. Macroscopic entanglement is a research field full of curiosity, and quantum optomechanical system is now considered to be useful for its investigation \cite{Vahala08,Girvin09,Woolley12,Chen13,Marquardt13,Tombesi02,Aspelmeyer07,Milburn11}. Generally speaking, the degree of entanglement is usually small (logarithmic negativity $E_{N}< 0.7$) near the zero temperature. It can not be obtained under room-temperature due to the stability conditions and the thermal noise of the mechanical modes.

Recently highly entangled quantum states (logarithmic negativity $E_{N}> 0.7$  near the zero temperature) in optomechanical system are discussed via various methods, such as cascaded cavity coupling \cite{Li13}, reservoir engineering \cite{Wang13}, Bogoliubov dark mode \cite{Tian13}, S{\o}rensen-M{\o}lmer approach \cite{HLWang13} and coherent feedback \cite{Clerk14}. The crucial component in these ideas is the generation of two-mode squeezing states. These results can play an important role in hybrid quantum networks, and can be extended to other parametrically coupled three-bosonic-mode systems, such as superconducting circuits coupled via Josephson junctions \cite{Girvin10}.

The dissipative ideas  in reservoir engineering  have been discussed and realized experimentally in atomic systems \cite{Huelga02,Cirac04,Cirac06,Plozik11,Cirac11}. In optomechanical system, a standard arrangement for generation of highly entangled state consists of two target modes (to be entangled but not directly coupled) and an intermediate mode (simultaneously coupled to the two target modes). Such systems can be used for quantum state transfer  \cite{Wang12,Tian12}. The dissipative environment of the two target modes can be controlled via reservoir engineering. Ultimately the two modes can be relaxed into an entangled state. This method can be realized with a high-frequency, low-Q mechanical resonator or coupling a high-Q mechanical mode to the third cavity mode. In Ref \cite{Wang13} the thermal noise of the mechanical mode was not directly discussed.

 In this paper steady-state entanglement in a four-mode optomechanical system is discussed, when the mechanical thermal noise is taken into account.
The situation that the two optomechanical couplings between the two target modes and the mechanical oscillator are equal can not be explained by the Bogoliubov-mode-based scheme \cite{Wang13}. This is connected with the ideas of quantum-mechanics-free subspace \cite{Caves12,Clerk13, Zhang13}. The mechanical thermal noise effect is discussed. Prominent steady-state entanglement can still exist under room-temperature.

\section{System}

We focus on a four-mode optomechanical system. The two target modes are two optical cavity modes, and the intermediate mode is a mechanical resonator, which is also coupled with a cooling cavity mode (see Fig. 1). When the frequency of the mechanical oscillator is much smaller than the frequency spacing of neighboring cavity modes, mainly the two target modes are coupled to the mechanical oscillator, and the mixing interaction between the two target modes and other excitation modes can be omitted \cite{Law95}. This single-mode expression is usually used in the discussion of optomechanical system, so the Hamiltonian of this system is
\begin{equation}
H=\frac{\omega_{m}}{2} (q^{2}+p^{2})+\sum_{i=1,2,3}(\omega_{i}a^{\dag}_{i}a_{i}+\sqrt{2}g_{i}a^{\dag}_{i}a_{i}q),
\end{equation}
where $a_{i}$ is the annihilation operators of cavity mode $i$, $q$ and $p$ are respectively the position operator and the momentum operator of the mechanical resonator, $g_{i}$ is the coupling strength between the cavity mode $i$ and the mechanical mode. Here 1,2 denote the two target modes, and 3 denotes the cooling mode. $\omega_{i}$ and $\kappa_{i}$ are the frequency and the damping rate of the cavity mode $i$.  $\omega_{m}$ and $\gamma_{m}$ are the frequency and the damping rate of the mechanical oscillator. The three cavity modes are respectively driven by lasers with frequencies $\omega_{1}-\omega_{m}$, $\omega_{2}+\omega_{m}$ and $\omega_{3}-\omega_{m}$.
Under the interaction picture with respect to the cavity drives, we can write $a_{i}=\bar{a}_{i}+d_{i}$,  $q=\bar{q}+\delta q$, $p=\bar{p}+\delta p$.  $\bar{a}_{i}$ is the stable cavity amplitude, and $\bar{q}$, $\bar{p}$ are the stationary mechanical position and momentum. If $|\bar{a}_{i}|\gg 1$, the optomechanical interaction can be linearized as follows
\begin{eqnarray}
H&=&\frac{\omega_{m}}{2}(\delta q^{2}+\delta p^{2})+\omega_{m} (d^{\dag}_{1}d_{1}-d^{\dag}_{2}d_{2}+d^{\dag}_{3}d_{3})  \nonumber\\
&&+\sum_{i=1,2,3}G_{i}(d^{\dag}_{i}+d_{i})\delta q,
\end{eqnarray}
where $G_{i}=\sqrt{2}g_{i}\bar{a}_{i}$.

\begin{figure}[tbp]
\includegraphics[width=1\columnwidth]{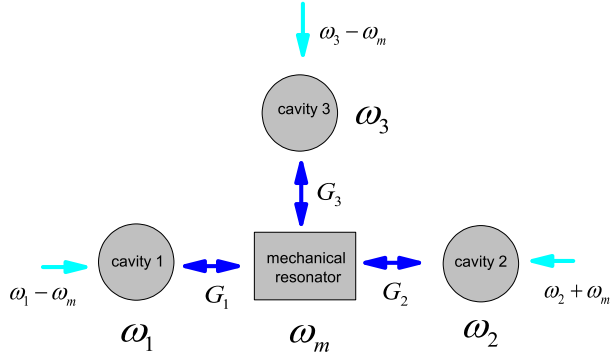}.
\caption{Schematic of a four-mode otpomechanical system, where three cavity modes are coupled to a single mechanical mode. 1,2 denote the two target modes to be entangled and 3 denotes the cooling mode which can be adiabatically eliminated when the dissipation rate $\kappa_{3}$ is large. Cavity 1,3 are driven at the red-detuned mechanical sideband, while cavity 2 is driven at the blue-detuned sideband.}
\end{figure}

The Heisenberg-Langevin equations for the linearized system become
\begin{eqnarray}
\delta \dot{q}&=&\omega_{m}\delta p,   \nonumber\\
\delta \dot{p}&=&-\omega_{m}\delta q-\gamma_{m}\delta p+G_{1}\delta X_{1}+G_{2}\delta X_{2}+G_{3}\delta X_{3}+\xi,   \nonumber\\
\delta \dot{X_{1}}&=&-\kappa_{1}\delta X_{1}+\omega_{m}\delta Y_{1}+\sqrt{2\kappa_{1}}X^{in}_{1},  \nonumber\\
\delta \dot{Y_{1}}&=&-\kappa_{1}\delta Y_{1}-\omega_{m}\delta X_{1}+G_{1}\delta q+\sqrt{2\kappa_{1}}Y^{in}_{1},  \nonumber\\
\delta \dot{X_{2}}&=&-\kappa_{2}\delta X_{2}-\omega_{m}\delta Y_{2}+\sqrt{2\kappa_{2}}X^{in}_{2},   \nonumber\\
\delta \dot{Y_{2}}&=&-\kappa_{2}\delta Y_{2}+\omega_{m}\delta X_{2}+G_{2}\delta q+\sqrt{2\kappa_{2}}Y^{in}_{2},   \nonumber\\
\delta \dot{X_{3}}&=&-\kappa_{3}\delta X_{3}+\omega_{m}\delta Y_{3}+\sqrt{2\kappa_{3}}X^{in}_{3},    \nonumber\\
\delta \dot{Y_{3}}&=&-\kappa_{3}\delta Y_{3}-\omega_{m}\delta X_{3}+G_{3}\delta q+\sqrt{2\kappa_{3}}Y^{in}_{3},
\end{eqnarray}
here the cavity field quadratures $\delta X_{i}\equiv (d_{i}+d^{\dag}_{i})/\sqrt{2}$ and  $\delta Y_{i}\equiv (d_{i}-d^{\dag}_{i})/i\sqrt{2}$
are defined. $\xi$ is the input noise operator. When the Q value of the mechanical oscillator is very high, $\langle \xi(t)\xi(t')+\xi(t')\xi(t)\rangle \simeq \gamma_{m}(2\bar{n}+1)\delta (t-t')$  can be satisfied \cite{Aspelmeyer07}, where $\bar{n}=(e^{\hbar \omega_{m}/k_{B}T}+1)^{-1}$ and $\hbar$, $k_{B}$ are the reduced Planck constant and the Boltzmann constant. $T$ is the bath temperature of the mechanical resonator.
$X^{in}_{i}$ and $Y^{in}_{i}$ are the input noise operators of the cavity mode $i (i=1,2,3)$ which are delta-correlated.

Equation (3) can be written in the following compact form
\begin{equation}
\dot{u}(t)=Au(t)+n(t),
\end{equation}
where
\begin{equation}
u^{T}=(\delta q,\delta p,\delta X_{1},\delta Y_{1},\delta X_{2},\delta Y_{2},\delta X_{3},\delta Y_{3}),
\end{equation}

\begin{eqnarray}
n^{T}=(0,\xi ,\sqrt{2\kappa_{1}}X^{in}_{1},\sqrt{2\kappa_{1}}Y^{in}_{1},\sqrt{2\kappa_{2}}X^{in}_{2},   \nonumber\\
\sqrt{2\kappa_{2}}Y^{in}_{2},\sqrt{2\kappa_{3}}X^{in}_{3},\sqrt{2\kappa_{3}}X^{in}_{3}),
\end{eqnarray}
and
\begin{scriptsize}
\begin{eqnarray}
A=\left(
\begin{array}{cccccccc}
 0 & \omega_{m} & 0 & 0 & 0 & 0  & 0 & 0 \\
 -\omega_{m} & -\gamma_{m} & G_{1} & 0 & G_{2} & 0  & G_{3} & 0 \\
  0 & 0 & -\kappa_{1} & \omega_{m} & 0  & 0  & 0 & 0 \\
   G_{1} & 0  & -\omega_{m} &  -\kappa_{1} & 0 & 0  & 0 & 0 \\
    0 & 0 & 0 & 0 &  -\kappa_{2} & -\omega_{m}  & 0 & 0 \\
     G_{2} & 0 & 0 & 0 & \omega_{m} &  -\kappa_{2}  & 0 & 0 \\
      0 & 0 & 0 & 0 & 0 & 0  &  -\kappa_{3} & \omega_{m} \\
       G_{3} & 0 & 0 & 0 & 0 & 0  & -\omega_{m} &  -\kappa_{3} \\
\end{array}
\right).
\end{eqnarray}
\end{scriptsize}
If all the eigenvlues of the matrix $A$ have negative real parts, the system is stable. The stability conditions can be obtained by use of the Routh-Hurwitz criteria \cite{Kaufman87}. Under the situation $\kappa_{1}=\kappa_{2}=\kappa$, the third cooling mode can be eliminated adiabatically, and the stability condition can be easily
expressed as \cite{Woolley12, Wang13}
\begin{equation}
G_{2}<G'=\sqrt{G_{1}^{2}+\frac{2\kappa}{\kappa_{3}}G_{3}^{2}}.
\end{equation}
This condition is necessary and sufficien. If $\kappa_{1}=\kappa_{2}$  and $G_{2}<G'$, the system is always stable. This condition is easily satisfied experimentally, which does not limit the coupling strength any more. When the system is stable, it reaches a steady four-mode Gaussian state, which can be fully characterized by a $8\times 8$ correlation matrix $V$ satisfying following equation
\begin{equation}
AV+VA^{T}=-D,
\end{equation}
where $D=$Diag$[0,\gamma_{m}(2\bar{n}+1),\kappa_{1},\kappa_{1},\kappa_{2},\kappa_{2},\kappa_{3},\kappa_{3}]$ is a diagonal matrix.

\section{Results}

To quantify the entanglement between the two target modes, we use the logarithmic negativity $E_{N}$. For two Gaussian modes $E_{N}$ can be calculated by the expression
\begin{equation}
E_{N}=\max\{0,-\ln2\eta^{-}\},
\end{equation}
where
\begin{equation}
\eta^{-}=\frac{1}{\sqrt{2}}\sqrt{\Sigma-\sqrt{\Sigma^{2}-4\det V_{ij}}}
\end{equation}
and
\begin{equation}
\Sigma=\det A+\det B-2\det C.
\end{equation}
The matrices $A$, $B$ and $C$ are $2\times 2$ blocks of the covariance matrix,
\begin{equation}
V_{ij}=\left(
\begin{array}{cc}
 A & C  \\
 C^{T} & B  \\
\end{array}
\right),
\end{equation}
where $i,j=1,2,3,m$ denote the red-sideband target mode, the blue-sideband target mode, the cooling mode and the mechanical oscillator. This condition is equivalent to Simon's partial transpose criterion \cite{Simon00}.

\begin{figure}[tbp]
\includegraphics[width=1.0\columnwidth]{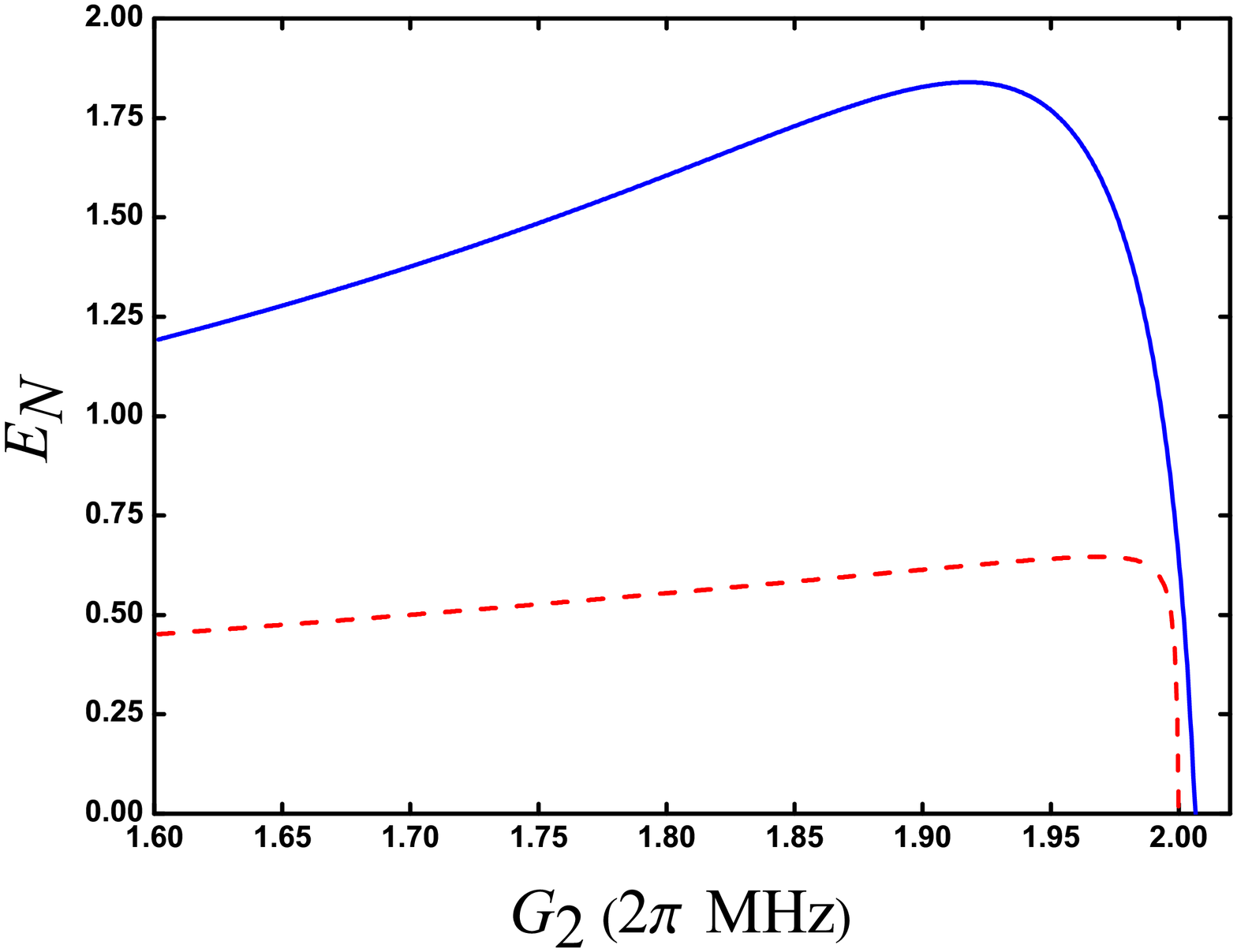}.
\caption{Stationary two target intracavity modes entanglement $E_{N}$ as a function of the interaction $G_{2}$. The parameters are chosen as $\omega_{m}=2\pi\times 10$ MHz, $G_{1}=2\pi\times 2$ MHz, $\kappa_{1}=\kappa_{2}=2\pi\times 0.02$ MHz, $\kappa_{3}=2\pi\times 0.5$ MHz, $\gamma_{m}=2\pi\times 100$ Hz, $T=300$ mK and $G_{3}=2\pi \times 0$ MHz (dashed-red line),  $2\pi \times 0.8$ MHz (solid-blue line).  }
\end{figure}

Fig. 2 shows the stationary entanglement $E_{N}$ between the two target intracavity modes as a function of the interaction $G_{2}$. We have taken parameters analogous to those of Ref. \cite{Wang13} that $\omega_{m}=2\pi\times 10$ MHz, $G_{1}=2\pi\times 2$ MHz, $\kappa_{1}=\kappa_{2}=2\pi\times 0.02$ MHz, $\kappa_{3}=2\pi\times 0.5$ MHz, $\gamma_{m}=2\pi\times 100$ Hz, $T=300$ mK and $G_{3}=2\pi \times 0$ MHz (dashed-red line), $2\pi \times 0.8$ MHz (solid-blue line). When $G_{3}=0$, the maximum entanglement approaches 0.7. However when $G_{3}=0.8$ MHz, the maximum entanglement is about 1.8, which is much larger than the usual value 0.7 induced by the two-modes squeezing interaction. So using the cooling mode can increase the entanglement significantly.

\begin{figure}[tbp]
\includegraphics[width=1.0\columnwidth]{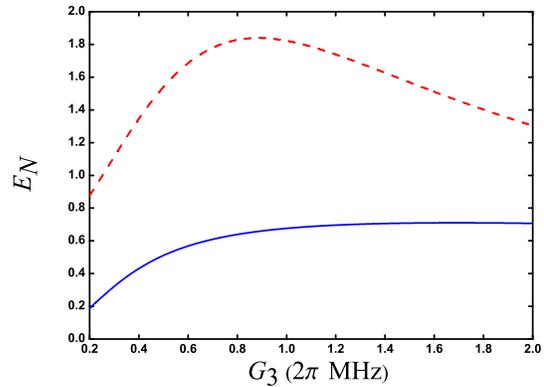}.
\caption{Stationary two target intracavity modes entanglement $E_{N}$ as a function of the interaction $G_{3}$. The parameters are chosen as $\omega_{m}=2\pi\times 10$ MHz, $G_{1}=2\pi\times 2$ MHz, $\kappa_{1}=\kappa_{2}=2\pi\times 0.02$ MHz, $\kappa_{3}=2\pi\times 0.5$ MHz, $\gamma_{m}=2\pi\times 100$ Hz, $T=300$ mK and $G_{2}=2\pi \times 1.9$ MHz (dashed-red line),  $2\pi \times 2$ MHz (solid-blue line). }
\end{figure}

The key finding in this paper is that, when $G_{2}=G_{1}$, the two target modes are still entangled ($E_{N}=0.6$).  We will show this result can not be explained via the Bogoliubov-mode-based scheme.
Two Bogoliubov modes for equation (3) by use of $\delta X=\frac{G_{1}X_{1}+G_{2}X_{2}}{\tilde{G}},\delta Y=\frac{G_{1}Y_{1}-G_{2}Y_{2}}{\tilde{G}}, \delta X'=\frac{G_{2}X_{1}+G_{1}X_{2}}{\tilde{G}}$ and  $\delta Y'=\frac{G_{2}Y_{1}-G_{1}Y_{2}}{\tilde{G}}$ can be introduced. We have
\begin{eqnarray}
\delta \dot{q}&=&\omega_{m}\delta p,  \nonumber\\
\delta \dot{p}&=&-\omega_{m}\delta q-\gamma_{m}\delta p+\tilde{G}\delta X+G_{3}\delta X_{3}+\xi, \nonumber \\
\delta \dot{X}&=&-\kappa \delta X+\omega_{m}\delta Y+\sqrt{2\kappa}X^{in},  \nonumber \\
\delta \dot{Y}&=&-\kappa \delta Y-\omega_{m}\delta X+\tilde{G}\delta q+\sqrt{2\kappa}Y^{in}, \nonumber \\
\delta \dot{X_{3}}&=&-\kappa_{3}\delta X_{3}+\omega_{m}\delta Y_{3}+\sqrt{2\kappa_{3}}X^{in}_{3},  \nonumber\\
\delta \dot{Y_{3}}&=&-\kappa_{3}\delta Y_{3}-\omega_{m}\delta X_{3}+G_{3}\delta q+\sqrt{2\kappa_{3}}Y^{in}_{3},
\end{eqnarray}
here $\tilde{G}=\sqrt{G_{1}^{2}-G_{2}^{2}}$, and $[\delta X,\delta Y]=[\delta X',\delta Y']=i$. It is obvious that the form can be valid only when $G_{2}<G_{1}$. This transformation is related to the Bogoliubov modes of the two target modes, and called Bogoliubov-mode-based scheme \cite{Wang13,HLWang13}. Here the mode $\delta X', \delta Y'$ is decoupled from the mechanical oscillator. When $G_{2}=G_{1}$, the form is not appropriate. From equation (14), if $G_{2}=G_{1}$, $\tilde{G}$ will be zero. The two target modes can completely decouple from the mechanical mode, and the entanglement will be zero. However, when $G_{3}> 0$, Fig.2 and Fig. 3 all show that the entanglement is not zero (the blue line in Fig. 3). When $G_{2}<G_{1}$ (dashed-red line), the two target modes can be greatly cooled. For a suitable $G_{3}$, we can have a strong steady-state entanglement 1.8 much larger than 0.7. When $G_{2}=G_{1}$, the entanglement also increases when $G_{3}$ increases, so the cooling mode can also be exploited to cool the two target modes.

\begin{figure}[tbp]
\includegraphics[width=1.0\columnwidth]{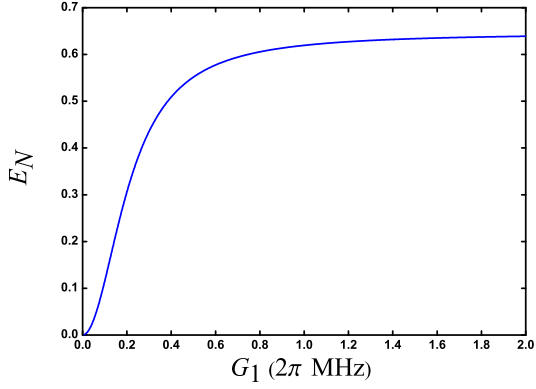}.
\caption{Stationary two target intracavity modes entanglement $E_{N}$ as a function of the interaction $G_{1}$. The parameters are chosen as $\omega_{m}=2\pi\times 10$ MHz, $G_{2}=G_{1}$, $\kappa_{1}=\kappa_{2}=2\pi\times 0.02$ MHz, $\gamma_{m}=2\pi\times 100$ Hz, $T=300$ mK and $G_{3}=2\pi \times 0.8$ MHz,  $\kappa_{3}=2\pi \times 0.5$ MHz. }
\end{figure}
\begin{figure}[tbp]
\includegraphics[width=1.0\columnwidth]{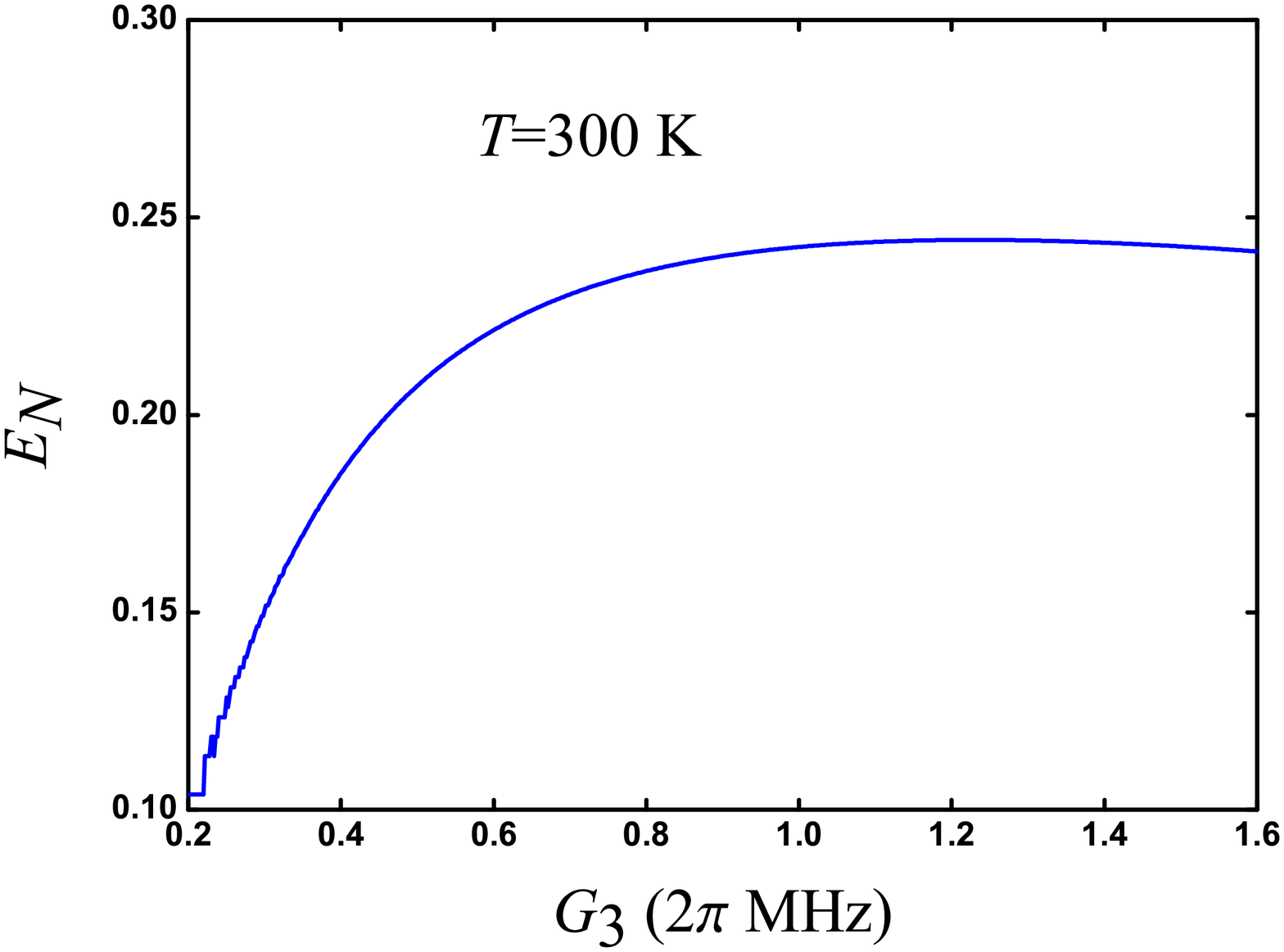}.
\caption{Stationary two target intracavity modes entanglement $E_{N}$ as a function of the interaction $G_{3}$ when $T=300$ K. The parameters are chosen as $\omega_{m}=2\pi\times 10$ MHz, $G_{1}=G_{2}=2\pi\times 2$ MHz, $\kappa_{1}=\kappa_{2}=2\pi\times 0.02$ MHz, $\kappa_{3}=2\pi\times 0.5$ MHz, $\gamma_{m}=2\pi\times 100$ Hz. }
\end{figure}
\begin{figure}[tbp]
\includegraphics[width=1.0\columnwidth]{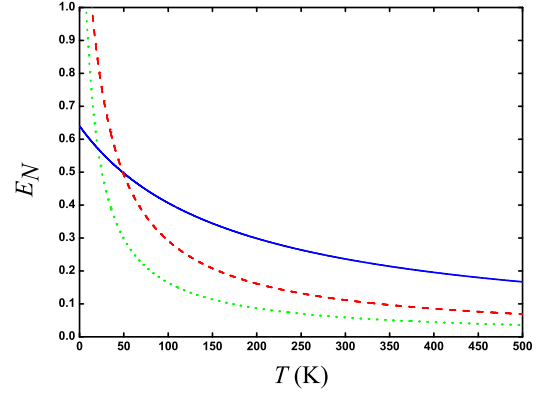}.
\caption{Stationary two target modes intracavity entanglement $E_{N}$ as a function of the mechanical thermal temprature $T$. The parameters are chosen as $\omega_{m}=2\pi\times 10$ MHz, $G_{1}=2\pi\times 2$ MHz,  $G_{3}=2\pi\times 0.8$ MHz, $\kappa_{1}=\kappa_{2}=2\pi\times 0.02$ MHz, $\kappa_{3}=2\pi \times 0.5$ MHz, $\gamma_{m}=2\pi\times 100$ Hz and $G_{2}=2\pi \times 1.9$ MHz (dotted-green line), $2\pi \times 1.95$ MHz (dashed-red line), $2\pi \times 2$ MHz (solid-blue line). }
\end{figure}

We notice that when $G_{2}=G_{1}=G$, equation (3) is connected with the idea of quantum-mechanics-free subspace  \cite{Caves12,Clerk13, Zhang13}. We can have a special form
\begin{eqnarray}
\delta \dot{q}&=&\omega_{m}\delta p,  \nonumber \\
\delta \dot{p}&=&-\omega_{m}\delta q-\gamma_{m}\delta p+\sqrt{2}G\delta X_{+}+G_{3}\delta X_{3}+\xi,\nonumber \\
\delta \dot{X_{+}}&=&-\kappa \delta X_{+}+\omega_{m}\delta Y_{-}+\sqrt{2\kappa}X_{+}^{in}, \nonumber  \\
\delta \dot{Y_{-}}&=&-\kappa \delta Y_{-}-\omega_{m}\delta X_{+}+\sqrt{2\kappa}Y_{-}^{in}, \nonumber \\
\delta \dot{X_{-}}&=&-\kappa \delta X_{-}+\omega_{m}\delta Y_{+}+\sqrt{2\kappa}X_{-}^{in}, \nonumber  \\
\delta \dot{Y_{+}}&=&-\kappa \delta Y_{+}-\omega_{m}\delta X_{-}+\sqrt{2}G\delta q+\sqrt{2\kappa}Y_{+}^{in}, \nonumber \\
\delta \dot{X_{3}}&=&-\kappa_{3}\delta X_{3}+\omega_{m}\delta Y_{3}+\sqrt{2\kappa_{3}}X^{in}_{3},  \nonumber\\
\delta \dot{Y_{3}}&=&-\kappa_{3}\delta Y_{3}-\omega_{m}\delta X_{3}+G_{3}\delta q+\sqrt{2\kappa_{3}}Y^{in}_{3}.
\end{eqnarray}
Here $\delta X_{+}=\frac{1}{\sqrt{2}}(\delta X_{1}+\delta X_{2})$, $\delta Y_{+}=\frac{1}{\sqrt{2}}(\delta Y_{1}+\delta Y_{2})$, $\delta X_{-}=\frac{1}{\sqrt{2}}(\delta X_{1}-\delta X_{2})$,  $\delta Y_{-}=\frac{1}{\sqrt{2}}(\delta Y_{1}-\delta Y_{2})$, and they satisfy the following relationships $[\delta X_{+}, \delta Y_{+}]=[\delta X_{-}, \delta Y_{-}]=i$ and $[\delta X_{+}, \delta Y_{-}]=[\delta X_{-}, \delta Y_{+}]=0$ which are EPR-like variables. When $G=0$, $\langle \delta X_{+}^{2}\rangle+\langle \delta Y_{-}^{2}\rangle=\langle \delta X_{-}^{2}\rangle+\langle \delta Y_{+}^{2}\rangle=1$. If $\langle \delta X_{+}^{2}\rangle+\langle \delta Y_{-}^{2}\rangle<1$ or $\langle \delta X_{-}^{2}\rangle+\langle \delta Y_{+}^{2}\rangle<1$, the two target modes can be entangled according to the criterion in \cite{Duan00}. This can be easily realized by adding the cooling mode. For equation (15), when $G>0$, $\delta X_{+}$ and $\delta Y_{-}$ is evaded from the mechanical oscillator, so $\langle \delta X_{+}^{2}\rangle+\langle \delta Y_{-}^{2}\rangle=1$ still holds. However $\delta Y_{+}$ and $\delta X_{-}$ will suffer from the oscillator. When the coupling strength between the cooling mode and the oscillator is large, the two target modes have a large dissipation. The dynamics of the oscillator can be eliminated adiabatically, and the two target modes are simultaneously generated or annihilated. Thus the entanglement between the two modes are created and $\langle \delta X_{-}^{2}\rangle+\langle \delta Y_{+}^{2}\rangle<1$ can be realized. This mechanism is very different from the previous Bogoliubov-mode-based scheme.

Fig. 4 plots the entanglement of the two target modes when $G_{2}=G_{1}$ and $T=300$ mK. When $G_{1}>2\pi\times0.8$ MHz, the entanglement can be about 0.6.
Fig. 5 presents the entanglement of the two target modes as a function of $G_{3}$ if $G_{2}=G_{1}$ and $T=300$ K (room temperature). When $G_{3}>2\pi\times0.4$ MHz, the entanglement is larger than 0.2, which is a prominent result (this result is impossible in previous investigations). Moreover Fig. 6 compares the two entanglement mechanism when the mechanical thermal noise is considered. It is shown that, at room-temperature and if $G_{2}=G_{1}$, we have the maximum entanglement 0.25 which can not be realized in the usual entanglement mechanism. This results will be important for hybrid quantum networks with optomechanical systems used under room temperature.

This paper is also inspired by the  S{\o}rensen-M{\o}lmer scheme in \cite{HLWang13} with the same interaction form, which has been discussed in \cite{Molmer99,Molmer00,Milburn99} to entangle trapped ions in a thermal environment, with which robust entanglement can be achieved. Ref. \cite{HLWang13} outlines a pulsed entanglement scheme in a three-modes optomechanical system featuring the S{\o}rensen-M{\o}lmer mechanism, which can generate strong entanglement in the weak and strong coupling regime. In contrast to the Bogoliubov-mode-based schemes, the S{\o}rensen-M{\o}lmer scheme are robust against the mechanical thermal noise, which has the same effect as this paper. However we notice that in the pulsed scheme the two target mode should detune from the mechanical oscillator, but here they have the same frequency.

\section{Conclusion}
In this papera four-mode optomechanical system is discussed in detail. The entanglement degree of the two target modes is calculated. When the two coupling strengths between the two target modes and the mechanical oscillator are equal, the result is related with the idea of quantum-mechanics-free subspace. Then   the Bogoliubov-mode-based scheme and the mechanism used in this paper are compared, when the mechanical thermal noise is considered. Importantly prominent entanglement of the two target modes still exist under room temperature. This results is important for the application of optomechanical networks.

\section{ACKNOWLEDGMENT}

This research is supported by National Natural Science Foundation of China (Grant No.11174109).

\end{document}